\documentclass{elsart3-1}

\usepackage{amssymb}

\usepackage[english,francais]{babel}

\newtheorem{e-proposition}[theorem]{Proposition}

\newtheorem{e-definition}[theorem]{Definition\rm}
\newtheorem{remark}{\it Remark\/}

\newtheorem{theoreme}{Th\'eor\`eme}[section]
\newtheorem{lemme}[theoreme]{Lemme}
\newtheorem{proposition}[theoreme]{Proposition}

\newtheorem{remarque}{\it Remarque}

\renewcommand{\theequation}{\arabic{equation}}
\def\og{\leavevmode\raise.3ex\hbox{$\scriptscriptstyle\langle\!\langle$~}}
\def\fg{\leavevmode\raise.3ex\hbox{~$\!\scriptscriptstyle\,\rangle\!\rangle$}}

\begin{document}
\centerline{Physique math\'{e}matique}
\begin{frontmatter}




%
\selectlanguage{francais}
\title{Probabilit\'{e}s et fluctuations quantiques}



\author[authorlabel1]{Michel FLIESS},
\ead{Michel.Fliess@polytechnique.edu}

\address[authorlabel1]{Projet ALIEN, INRIA Futurs \\ \& \'Equipe MAX, LIX (CNRS, UMR 7161),
\'Ecole polytechnique, 91128 Palaiseau, France}


\medskip
\selectlanguage{francais}
\begin{center}
{\small Re\c{c}u le *****~; accept\'e apr\`es r\'evision le +++++\\
Pr\'esent\'e par {\pounds}{\pounds}{\pounds}{\pounds}{\pounds}}
\end{center}

\begin{abstract}
\selectlanguage{francais}
Cette Note esquisse une construction math\'{e}matique simple et naturelle
du caract\`{e}re probabiliste de la m\'{e}canique quantique. Elle utilise
l'analyse non standard et repose sur
l'interpr\'{e}tation due \`{a} Feynman, mise en avant dans certaines approches fractales,
du principe d'incertitude de Heisenberg, c'est-\`{a}-dire des fluctuations
quantiques.
On aboutit ainsi \`{a} des \'{e}quations diff\'{e}rentielles
stochastiques, comme dans la m\'{e}canique stochastique de Nelson,
d\'{e}coulant de marches al\'{e}atoires infinit\'{e}simales. {\it Pour citer cet article: M.
Fliess, C. R. Acad. Sci. Paris, Ser. I 344 (2007).} \vskip
0.5\baselineskip

\selectlanguage{english} \noindent{\bf Abstract} \vskip
0.5\baselineskip \noindent {\bf Probabilities and quantum fluctuations.}
This Note is sketching a simple and natural mathematical construction
for explaining the probabilistic nature of quantum mechanics.
It employs nonstandard analysis and is based on
Feynman's interpretation of the Heisenberg uncertainty principle,
i.e., of the quantum fluctuations, which was brought to the
forefront in some fractal approaches. It results, as in Nelson's
stochastic mechanics, in stochastic differential equations which are
deduced from infinitesimal random walks.
 {\it To cite
this article: M. Fliess, C. R. Acad. Sci. Paris, Ser. I 344 (2007).}
\end{abstract}
\end{frontmatter}

\selectlanguage{english}
\section*{Abridged English version}

We are sketching a ``simple and natural'' explanation of the probabilistic
nature of quantum mechanics. This is achieved by utilizing the mathematical
formalism of nonstandard analysis and Feynman's interpretation of the
Heisenberg uncertainty principle, i.e., of the quantum fluctuations.

\noindent{\bf Non-technical presentation of the main ideas.}
This summary is intended for readers who are not familiar
with non-standard analysis.
As often in nonstandard analysis (see, e.g., \cite{al,nelson-proba}), we replace
a continuous time interval by an infinite ``discrete'' set of
infinitely closed time instants.
Substituting $m \frac{\Delta x}{\Delta t}$ to $\Delta p$ in the well
known expression of the uncertainty principle, where $x$
is the position, $m$ the mass, $p$ the momentum, $h = 2 \pi \hbar$
the Planck constant, yields equation (\ref{rhei}). We rewrite it by
postulating that the quantity (\ref{fond}), where $\delta t
> 0$ is a given infinitesimal, is limited and appreciable, i.e., it is
neither infinitely large nor infinitely small.
Those computations are stemming from Feynman's interpretation
\cite{feynman0,feynman} of the uncertainty principle (see, also,
\cite{kr,not1}): The ``quantum trajectories'' are
fractal curves, of Hausdorff dimension $2$.
``Weak'' mathematical
assumptions permit to derive the infinitesimal difference equation
(\ref{ma}).
The lack of any further physical assumption yields
the equiprobability of $+1$ and $- 1$.
If $x$ is {\em Markov}, i.e., if $b$ and $\sigma$ are
are functions of $t$ and $x(t)$, and not of $\{ x(\tau) | 0 \leq
\tau < t \}$, the corresponding infinitesimal random walk is \og
equivalent\fg ~ to a stochastic differential equation in the usual
sense (see, e.g., \cite{al,benoit}).

\begin{remark}
More or less analogous random walks have already been introduced in
the literature (see, e.g., \cite{badiali,gud,ord1,ord2}), but in another
context.
\end{remark}

\noindent{\bf Non-standard analysis.}
Replace the interval $[0, 1]$ by the set ${\mathfrak{Q}} = \{k
\delta t \mid 0 \leq k \leq N_q \}$, $\delta t = \frac{1}{N_q}$,
where $N_q$ is an unlimited integer.
A function $x: {\mathfrak{Q}} \rightarrow {\mathbb{R}}$ is said to
verify the {\em Heisenberg condition} if, and only if, for any $t
\in {\mathfrak{Q}} \setminus \{1\}$, equation (\ref{heis}) is
satisfied, where $\sigma^{\star}_{t} \in {\mathbb{R}}$,
$\sigma^{\star}_{t}> 0$, is limited and appreciable.
The lack of any further physical assumption leads us to postulate
the following properties for $\varepsilon (t)$ in equation
(\ref{heis-prob}):
the random variables $\varepsilon (t)$, $t \in {\mathfrak{Q}} \setminus \{1\}$,
are independent; $\mbox{\rm Pr}
(\varepsilon (t) = +1) = \mbox{\rm Pr} (\varepsilon (t) = -1) =
\frac{1}{2}$.
The next properties are ``natural'' for the stochastic processes
$\varepsilon$ and $x$ (see \cite{nelson-proba} and
\cite{benoit0,benoit}):
the random variables $x(0)$ and $\varepsilon (t)$, $t \in {\mathfrak{Q}}$,
are independent;
$x(t)$, $t > 0$, is function of $\{\varepsilon (\tau)| 0 \leq \tau \leq
t - \delta t \}$.
Set $E^{\varepsilon}_{t} (\bullet) = E_{t} (\bullet | \varepsilon
(0), \varepsilon (\delta t), \dots, \varepsilon (t - \delta t))$.
Equation (\ref{qf}) yields the decomposition (\ref{decomp})
\cite{nelson-proba}, where $\eta$ is a stochastic process, such that
$E^{\varepsilon}_{t} (\eta (t)) = 0$, $E^{\varepsilon}_{t} ((\eta
(t))^2) = 1$. Then, for all $t \in {\mathfrak{Q}} \setminus \{1\}$,
$\sigma^{\star}_{t} \simeq s^{\star}_{t}$, $\varepsilon (t) =
\eta (t)$.

We will not make any distinction between two equations of type
(\ref{decomp}) if the coefficients are limited and infinitely
closed. The process $x$ is said to be {\em Markov}, or to satisfy
the {\em Markov condition}, if, and only if, there exist $b \simeq
D^\varepsilon x (t)$, $\sigma \simeq s^{\star}_{t}$, such that
they are functions of $t$ et $x(t)$, and not of $\{x(\tau) |  0 \leq
\tau < t \}$.
Assume that $x$ is Markov. Consider the infinitesimal random walk
(\ref{mad}). According to \cite{benoit0} (see, also, \cite{benoit})
it defines a diffusion if, and only if, the following conditions are
satisfied (see \cite{al} for another approach): $b$ and $\sigma$ are of class $S^0$;
the shadows $\tilde{b}$ and $\tilde{\sigma}$ of
$b$ et $\sigma$ are of class $C^\infty$; the function $\tilde{\sigma}$ is always strictly
positive; the random variable $x(0)$ is almost surely limited.

\setcounter{section}{0} \selectlanguage{francais}
\section{Introduction}
\label{intro}
Le caract\`{e}re probabiliste du quantique, confirm\'{e} par
tant d'exp\'{e}riences, proc\`{e}de, depuis Born, Bohr et
l'\'{e}cole dite de Copenhague-G\"{o}ttingen,
d'une axiomatique de la fonction d'onde de Schr\"odinger, qui,
comme on le sait,
a fait na\^{\i}tre, \`{a} cause de son \'{e}tranget\'{e} et de ses apparents paradoxes,
d'innombrables commentaires (cf. \cite{bitbol}).
Cette Note (voir \cite{prel} pour une version pr\'{e}liminaire) vise \`{a} d\'{e}duire
cette nature
al\'{e}atoire de consid\'{e}rations
math\'{e}matiques \og simples et naturelles \fg. Elle repose sur
\begin{itemize}
\item l'interpr\'{e}tation par Feynman \cite{feynman0,feynman} du
principe d'incertitude de Heisenberg, c'est-\`{a}-dire des fluctuations quantiques;

\item l'{\em analyse non standard} de Robinson \cite{robinson}, d\'{e}j\`{a}
employ\'{e}e en quantique (voir, par exemple, \cite{al}, sa
bibliographie, et \cite{gud,not1}), et, plus pr\'{e}cis\'{e}ment, \begin{itemize}
\item une
discr\'{e}tisation infinit\'{e}simale du temps,
\item le calcul non standard des probabilit\'{e}s
de Nelson \cite{nelson-proba} et son extension aux \'{e}quations
diff\'{e}rentielles stochastiques \cite{benoit0,benoit}.
\end{itemize}
\end{itemize}
On aboutit \`{a} des \'{e}quations diff\'{e}rentielles stochastiques gr\^ace \`{a} de
marches al\'{e}atoires infinit\'{e}simales. Nous empruntons ainsi \`{a} des degr\'{e}s divers
\`{a} trois alternatives aux fondements du quantique,
parfois tr\`{e}s controvers\'{e}es, dont les liens avaient d\'{e}j\`{a} \'{e}t\'{e} constat\'{e}s dans la
litt\'{e}rature (cf. \cite{badiali}): l'approche fractale, la m\'{e}canique stochastique
et les marches al\'{e}atoires.
\begin{enumerate}
\item Nous suivons la premi\`{e}re, port\'{e}e par
Nottale \cite{not1} (voir aussi \cite{kr}), dans sa mise en valeur de
l'interpr\'{e}tation due \`{a}
Feynman du principe d'incertitude, qui semblait plus ou moins oubli\'{e}e.
\item Les \'{e}quations diff\'{e}rentielles stochastiques ordinaires sont un pivot
de la seconde, invent\'{e}e
par F\'enyes \cite{fen},
d\'{e}velopp\'{e}e et popularis\'{e}e par
Nelson \cite{physrev,nelson-brown,nelson-quant}. Cette m\'{e}canique
continue \`{a} susciter
une litt\'{e}rature abondante (cf.
\cite{blanchard,davidson,fri,oron,pavon,pen,smolin,smolin2}),
en d\'{e}pit de critiques virulentes (cf. \cite{chung,fri}).
Voir, par exemple,
\cite{ayz,chung} pour une comparaison avec la m\'{e}canique quantique {\em euclidienne},
que beaucoup, Nelson y compris, jugent plus convaincante.
\item Renvoyons pour la troisi\`{e}me aux fameux \'{e}chiqier ({\em chessboard})
de Feynman \cite{feynman}, aux travaux d'Ord (cf. \cite{ord1,ord2})
et \`{a} \cite{badiali,gud}.
\end{enumerate}

\begin{remarque}On se restreint \`{a}
un espace de dimension $1$ afin d'all\'{e}ger l'\'{e}criture.
\end{remarque}

\begin{remarque}Alors que certains travaux autour de la m\'{e}canique
stochastique se rattachent \`{a} l'automatique et, plus pr\'{e}cis\'{e}ment, \`{a}
la commande optimale stochastique (cf. \cite{guerra}), notre point
de vue fait suite \`{a} une approche nouvelle \cite{fliess} de
l'estimation et de l'identification, en automatique et signal.
Rappelons les liens entre le quantique actuel et certains aspects du
traitement de signal (cf. \cite{flandrin}).
\end{remarque}

\begin{remarque}On s'efforcera de d\'{e}gager les
rapports avec des travaux actuels sur le plongement stochastique
\cite{cresson1,cresson2}.
\end{remarque}

\section{Pr\'{e}sentation g\'{e}n\'{e}rale}
Voici une synth\`{e}se des principales id\'{e}es directrices, sans
pr\'{e}tention de rigueur, mais, esp\'{e}rons-le, claire et transparente,
destin\'{e}e aux lecteurs peu au fait de l'analyse non standard:

\begin{enumerate}

\item On remplace, comme souvent en non-standard (cf. \cite{al,nelson-proba}), le temps
continu par un ensemble infini \og discret \fg ~ d'instants
infiniment proches.

\item
Les calculs suivants \'{e}manent d'une reformulation du principe
d'incertitude, due \`{a} Feynman \cite{feynman0,feynman} (voir, aussi,
\cite{kr,not1}):
la \og trajectoire quantique \fg ~ est une courbe
continue, non d\'{e}rivable, c'est-\`{a}-dire fractale, de dimension de
Hausdorff $2$, traduisant la \og {\em Zitterbewegung} \fg, c'est-\`{a}-dire
les fluctuations quantiques. Soient $x$ la position, $m$ la
masse, $p$ la quantit\'{e} de mouvement,
$h = 2 \pi \hbar$ la constante de Planck. Le symbole $\Delta$
d\'{e}signe un \og petit \fg ~ accroissement; ainsi $\Delta x = x(t + \Delta t) - x(t)$
et, donc, $\Delta p \simeq m \frac{\Delta x}{\Delta t}$.
L'expression famili\`{e}re $\Delta x \Delta p \gtrsim \hbar$
du principe d'incertitude devient (cf. \cite{kr}, p. 85):
\begin{equation}\label{rhei}
\frac{(\Delta x)^2}{\Delta t} \gtrsim \frac{\hbar}{m}
\end{equation}
Soit $\delta t > 0$
infinit\'{e}simal donn\'{e}. On r\'{e}\'{e}crit (\ref{rhei}) en postulant que
\begin{equation}\label{fond}\frac{ \left( x(t + \delta t) - x(t)
\right)^2}{\delta t}\end{equation} est limit\'{e} et appr\'{e}ciable, c'est-\`{a}-dire ni
infiniment grand ni infiniment petit.

\begin{remarque}
Voir, \`{a} propos de l'irr\'{e}versibilit\'{e} thermodynamique,
\cite{badiali} pour une condition voisine de (\ref{rhei})-(\ref{fond}) dans un
espace-temps discret.
\end{remarque}

\item On en d\'{e}duit, moyennant des hypoth\`{e}ses math\'{e}matiques \og faibles \fg,
l'\'{e}quation aux diff\'{e}rences infinit\'{e}simales
\begin{equation}\label{ma}
x(t + \delta t) = x(t) + b \delta t \pm \sigma \sqrt{\delta t}
\end{equation}

\item L'absence de toute hypoth\`{e}se physique suppl\'{e}mentaire conduit \`{a}
postuler l'\'{e}quiprobabilit\'{e} de $\pm 1$.

\item Si $x$ est {\em markovien},
c'est-\`{a}-dire si $b$ et $\sigma$ sont fonctions de $t$ et $x(t)$, et
non de $\{ x(\tau) | 0 \leq \tau < t \}$, la marche al\'{e}atoire
infinit\'{e}simale d\'{e}finie par (\ref{ma}) \og \'{e}quivaut \fg ~ \`{a} une \'{e}quation diff\'{e}rentielle
stochastique au sens usuel (cf. \cite{al,benoit0,benoit}), comme en m\'{e}canique
stochastique.

\end{enumerate}

\section{Analyse non standard}
On r\'{e}dige selon la terminologie {\it IST} de
\cite{nelson-ist,nelson-proba}.

\subsection{Condition de Heisenberg}\label{cond}
Rempla\c{c}ons l'intervalle $[0, 1]$ par ${\mathfrak{Q}} = \{k \delta t
\mid 0 \leq k \leq N_q \}$, $\delta t = \frac{1}{N_q}$, o\`u $N_q$
est un entier illimit\'{e}. Appelons ${\mathfrak{Q}}$ et $\delta t$,
respectivement, l'{\em \'{e}chelle} et le {\em pas de temps quantiques}.
Une fonction $x: {\mathfrak{Q}} \rightarrow {\mathbb{R}}$ est dite
v\'{e}rifier la {\em condition de Heisenberg} si, et seulement si, pour
tout $t \in {\mathfrak{Q}} \setminus \{1\}$,

\begin{equation}
\label{heis} \frac{ \left( x(t + \delta t) - x(t) \right)^2}{\delta
t} \simeq ({\sigma}^{\star}_{t})^2 \end{equation} o\`{u}
${\sigma}^{\star}_{t} \in {\mathbb{R}}$, ${\sigma}^{\star}_{t} > 0$,
est limit\'{e} et appr\'{e}ciable. Il en d\'{e}coule

\begin{equation}
\label{heis-prob}  x(t + \delta t) - x(t) \simeq \varepsilon
(t)({\sigma}^{\star}_{t})\sqrt{\delta t} \quad \mbox{\rm o\`{u}} ~
\varepsilon (t) = \pm 1
\end{equation}

\subsection{\'Equiprobabilit\'{e} de $+1$ et $-1$}
Postulons les deux propri\'{e}t\'{e}s suivantes, qui d\'{e}coulent \og
naturellement \fg ~ de l'absence de toute hypoth\`{e}se physique
suppl\'{e}mentaire:
\begin{itemize}
\item les $\varepsilon (t)$, $t \in {\mathfrak{Q}} \setminus \{1\}$, sont des
variables al\'{e}atoires ind\'{e}pendantes,
\item $\mbox{\rm Pr}
(\varepsilon (t) = +1) = \mbox{\rm Pr} (\varepsilon (t) = -1) =
\frac{1}{2}$.
\end{itemize}

\subsection{Processus stochastiques}
Comme $\varepsilon$ peut \^etre vu comme un processus stochastique,
il est loisible de postuler les propri\'{e}t\'{e}s suivantes (cf.
\cite{nelson-proba} et \cite{benoit0,benoit}):
\begin{itemize} \item $x(t)$ est un processus stochastique;
\item les variables al\'{e}atoires $x(0)$ et $\varepsilon (t)$, $t \in
{\mathfrak{Q}}$, sont ind\'{e}pendantes;
\item $x(t)$, $t > 0$, est fonction de $\{\varepsilon (\tau)| 0 \leq \tau \leq
t - \delta t \}$.
\end{itemize}
Notons $E^{\varepsilon}_{t} (\bullet)$ l'esp\'{e}rance conditionnelle du
processus $\bullet$ en l'instant $t \in {\mathfrak{Q}}$, sachant le
pass\'{e} du processus $\varepsilon$. Posons
\begin{equation}\label{qf}
\begin{array}{c}
D^\varepsilon x(t) = \frac{E^{\varepsilon}_{t} (x(t + \delta
t) - x(t))}{\delta t}  \\
s^{\star}_{t} \simeq \sqrt{\frac{(x(t + \delta t) - x(t) -
D^\varepsilon x(t) \delta t)^2}{\delta t}}
 \end{array}
\end{equation}
Supposons $D^\varepsilon x(t)$ limit\'{e}. On obtient la d\'{e}composition \cite{nelson-proba} (voir,
aussi, \cite{benoit0}):
\begin{equation}\label{decomp}
x (t + \delta t) = x(t) + (D^\varepsilon x(t)) \delta t +
(s^{\star}_{t}) \eta (t) \sqrt{\delta t}
\end{equation}
o\`u $\eta$ est un processus stochastique, tel que
$E^{\varepsilon}_{t} (\eta (t)) = 0$, $E^{\varepsilon}_{t} ((\eta
(t))^2) = 1$. La proposition suivante relie les quantit\'{e}s
$\sigma^{\star}_{t}$ et $\varepsilon (t)$ des paragraphes pr\'{e}c\'{e}dents
aux nouvelles:

\begin{proposition}
Pour tout $t \in {\mathfrak{Q}} \setminus \{1\}$, il est loisible de poser
$\sigma^{\star}_{t} \simeq s^{\star}_{t}$, $\varepsilon (t) =
\eta (t)$.
\end{proposition}

\subsection{Condition de Markov}
Soit
$$
x_i (t + \delta t) = x_i(t) + (D^\varepsilon x_i(t)) \delta t +
(s^{\star}_{i, t}) \varepsilon (t) \sqrt{\delta t}
$$
o\`u  $i = 1, 2$. La propri\'{e}t\'{e} suivante est facile:
\begin{lemme}
Si, pour tout $t \in {\mathfrak{Q}} \setminus \{1\}$, les quantit\'{e}s
infiniment proches $D^\varepsilon x_1(t) \simeq D^\varepsilon
x_2(t)$, $s^{\star}_{1, t} \simeq s^{\star}_{2, t}$,
$x_1(0) \simeq x_2 (0)$ sont limit\'{e}es en valeur absolue par $C \in
\mathbb{R}$, la diff\'{e}rence $x_1 (t) - x_2 (t)$ reste infinit\'{e}simale
pour tout $t \in \mathfrak{Q}$.
\end{lemme}
Nous ne distinguerons pas deux \'{e}quations de type (\ref{decomp}) avec
coefficients limit\'{e}s et infiniment proches. On dit que $x$ est {\em
markovien}, ou satisfait la {\em condition de Markov}, si, et
seulement si, il existe des quantit\'{e}s $b \simeq D^\varepsilon x
(t)$, $\sigma \simeq s^{\star}_{t}$ fonctions de $t$ et
$x(t)$, et non de $\{x(\tau) |  0 \leq \tau < t \}$.

\subsection{Diffusions}
Supposons, dor\'{e}navant, $x$ markovien et consid\'{e}rons la marche
al\'{e}atoire infinit\'{e}simale
\begin{equation}
\label{mad} x(t + \delta t) = x(t) + b(t, x(t)) \delta t + \sigma
(t, x(t)) \varepsilon (t) \sqrt{\delta t}
\end{equation} Renvoyons \`{a} \cite{nelson-proba} (voir, aussi,
\cite{benoit0,benoit})
pour la notion d'{\em \'{e}quivalence} de processus stochastiques. Le
repr\'{e}sentant (\ref{mad}) d\'{e}finit, selon \cite{benoit0} (voir, aussi,
\cite{benoit}), une diffusion si, et seulement si, les conditions
suivantes sont satisfaites:
\begin{itemize}
\item $b$ et $\sigma$ sont de classe $S^0$,

\item les ombres $\tilde{b}$ et $\tilde{\sigma}$ de
$b$ et $\sigma$ sont de classe $C^\infty$,
\item la fonction $\tilde{\sigma}$ est toujours strictement
positive,
\item la variable al\'{e}atoire $x(0)$ est presque s\^urement limit\'{e}e.
\end{itemize}

\begin{remarque}
Voir \cite{al} pour une autre approche des liens entre (\ref{mad})
et \'{e}quations diff\'{e}rentielles stochastiques, o\`u les conditions
requises pour $b$ et $\sigma$ sont sensiblement diff\'{e}rentes.
\end{remarque}

\vspace{0.1cm} \noindent{\bf Remerciements}. L'auteur exprime sa
reconnaissace \`a E. Beno\^{\i}t (La Rochelle) pour lui avoir communiqu\'{e}
la r\'{e}f\'{e}rence \cite{benoit0}, \`{a} J.-M. L\'{e}vy-Leblond (Nice) pour des
remises en question salutaires, \`{a} P. Rouchon (Paris) pour des
discussions pr\'{e}cieuses, et \`{a} T. Sari (Mulhouse) pour des conseils utiles.


\end{document}